# Quasi-real-time dual-comb spectroscopy with 750-MHz Yb:fiber combs

**HAOCHEN TIAN,**[1,2] **RUNMIN LI,**[1] **LUKASZ A. STERCZEWSKI,**[3] **TAKASHI KATO,**[1,4] **AKIFUMI ASAHARA,**[1] **AND KAORU MINOSHIMA**[1,*]

[1]*Graduate School of Informatics and Engineering, The University of Electro-Communications, 1-5-1 Chofugaoka, Chofu, Tokyo 182-8585, Japan*
[2]*JSPS Postdoctoral Fellowships for Research in Japan*
[3]*Faculty of Electronics, Photonics and Microsystems, Wroclaw University of Science and Technology, Wybrzeże Wyspiańskiego 27, 50-370 Wrocław, Poland*
[4]*PRESTO, JST, 1-5-1 Chofugaoka, Chofu, Tokyo, Japan*
*\*k.minoshima@uec.ac.jp*

**Abstract:** We present quasi-real-time dual-comb spectroscopy (DCS) using two Yb:fiber combs with ~750 MHz repetition rates. A computational coherent averaging technique is employed to correct timing and phase fluctuations of the measured dual-comb interferogram (IGM). Quasi-real-time phase correction of 1-ms long acquisitions occurs every 1.5 seconds and is assisted by coarse radio frequency (RF) phase-locking of an isolated RF comb mode. After resampling and global offset phase correction, the RF comb linewidth is reduced from 200 kHz to ~1 kHz, while the line-to-floor ratio increases 13 dB in power in 1 ms. Using simultaneous offset frequency correction in opposite phases, we correct the aliased RF spectrum spanning three Nyquist zones, which yields an optical coverage of ~180 GHz around 1.035 µm probed on a sub-microsecond timescale. The absorption profile of gaseous acetylene is observed to validate the presented technique.

DOI: https://doi.org/10.1364/OE.460720

## 1. Introduction

Over two decades since the first demonstration of dual-comb spectroscopy (DCS) [1], this powerful technique has been widely applied to gaseous species monitoring [2-4]. Many modalities have also extended the original DCS concept like ultrafast time-resolved measurements [5], photoacoustic measurements [6], electro-optic sampling probing the electric field [7,8], or hyperspectral holography [9].

In DCS measurements, using optical frequency combs (OFCs) with ~ GHz level comb spacing as laser sources which enables high update rate and broad Nyquist range has been demonstrated by several research groups [10-15]. Nevertheless, precise spectroscopy at prolonged integration times relies on mutual coherence between the two OFCs, which conventionally requires carefully intrinsic noise suppression in lasers along with tight phase locking [16]. This is always not easy to be implemented in high-repetition-rate combs (~ GHz repetition rate), owing to the following two reasons: (i) the intrinsic phase noise is higher than conventional combs (~100 MHz repetition rate) [17]. (ii) obtaining carrier-envelope offset frequency beat with sufficient SNR requires more efforts because the pulse energy is generally not enough to get octave-spanning spectrum. Consequently, DCS measurement using free-running high-repetition-rate comb with data post-processing can be an attractive solution. Phase fluctuations in the interferogram (IGM) signal, which originate from relative phase fluctuations between the two free-running sources can be digitally corrected [18]. This is possible because the IGM's relative phase noise degrades the spectroscopic measurement far more than the loss of precision due to MHz-level drifts in the absolute frequency axis of the sample-interrogating comb. Arguably, it is the main motivation for developing mode-resolved free-running dual-comb spectrometers: the linewidth of the free-running optical frequency combs (~ MHz level) is orders of magnitude narrower than that required for measurements of gaseous species at atmospheric pressure (~ GHz level linewidths).

In the digital correction approach, the correction signals can be obtained from optical beats between the two combs and two intermediate free-running continuous wave (CW) lasers [18]. This concept was extended by Zhu *et al*, who demonstrated digital phase correction using only one free-running CW laser as an optical intermediary [19]. After correction, a sub-hertz relative linewidth, ~1 ns relative timing jitter, and 0.2 rad precision in the carrier phase were achieved. Phase noise correction can be also realized without any intermediary oscillators, as reported to date for fiber lasers [20-21] and chip-scale sources [22-23]. This approach is highly advantageous in spectral regions different from the mature 1.5 μm telecom band, where the availability of additional lasers or photodetectors is limited, like the mid-infrared or terahertz. The main requirement is sufficient short-term stability between the combs: on a $1/\Delta f_{rep}$ timescale the relative phase must not drift by more than π, where $\Delta f_{rep}$ is the repetition rate difference [24]. This condition is often fulfilled for free-running DCS systems with MHz-level RF line spacing or much denser single-cavity dual-comb lasers [4]. Whereas virtually all computational correction techniques perform well with non-aliased spectra (located between DC and half the comb's repetition rate), scenarios where the spectrum spans over more than half the comb's repetition rate $f_{rep}$ (first Nyquist zone) become a challenge. This simply relates to the distortion of the IGM and the completely different signal model in the case of Kalman filter estimation [22]. Although aliasing may be seen as a violation of the critical Nyquist–Shannon sampling theorem that often leads to spectral clutter and distortion of spectroscopic information, some scenarios take advantage of it. For instance, when aliased ($f_{rep}/2$- or DC-reflected) and non-aliased comb lines do not spectrally overlap, one can double the number of comb lines on a given photodetector bandwidth via spectral interleaving, as proposed by Schiller [1] or by arbitrary factors using subsampling [25]. This is particularly important for low-bandwidth photodetectors or DCS systems with a large repetition rate difference $\Delta f_{rep}$ for time-resolved sub-microsecond resolution studies. Spectra employing controlled aliasing have groups of RF comb lines located in different Nyquist zones spaced with intervals $f_{rep}/2$, which move in counter phase when perturbed by phase noise.

Previously, a technique called CoCoA (Computational Coherent Averaging) algorithm was proposed, where correcting the in-phase and counter phase noise of comb modes located in different Nyquist zones can be realized. It is because, in this technique, it can extract the relative offset frequency $\Delta f_0$ from a single comb line moving in any direction while extracting the repetition rate from all lines (even those aliased) [26]. This technique can be perfectly tailored for OFCs with GHz-level $f_{rep}$ and MHz-level $\Delta f_{rep}$, because the RF comb modes are well-isolated in RF domain and the amount of power per comb tooth is increased with enlarged Nyquist frequency (aliasing-free optical bandwidth).

However, despite their theoretical hardware realization potential, prior demonstrations of computational-only phase correction involved manual data post-processing. Rapid developments in software (data processing techniques) and hardware (digitizer or field-programmable gate array) have pushed DCS towards high-speed and real-time measurements [27-29]. In particular, the adaptive interferogram sampling technique presented by Ideguchi *et al* enables real-time DCS with free-running OFCs across different spectral regions and with different comb platforms [30-33], but at the expense of a significant complication on the hardware side, as it requires additional frequency multipliers, photodetectors, CW lasers, and dedicated digitizers.

In this work, we develop quasi-real-time dual-comb spectroscopy technique utilizing two free-running 750-MHz Yb:fiber combs assisted with RF comb line phase locking. 1-ms-long sequences of IGMs ($10^6$ samples) are phase-corrected by CoCoA technique in every 1.5 seconds and averaged in the spectral domain. The 750-MHz repetition rate combs with ~MHz level $\Delta f_{rep}$ permit the RF comb modes to be easily isolated and used for extracting the correction signals. To ensure this capability over longer timescales, one of the RF comb lines at 70 MHz is phase-locked to an RF reference through pump power feedback. The simple yet robust phase-locking prevents the comb modes from drifting, which allows the data sampling and phase correction in a quasi-real-time manner. As a demonstration, we show power-averaging of 20 phase-corrected tooth-resolved 1-ms long spectra acquired over more than 30 seconds. The short-term RF comb mode linewidth reduces from 200 kHz to the Fourier limit of ~1 kHz at 1 ms. Because the coarse lock increases the RF mode stability, it further enables RF spectral interleaving in one Nyquist zone, as a result, simultaneous correction of the offset frequency in-phase and counter phase recovers the RF spectrum spanning 3 Nyquist zones (from DC to $3f_{rep}/2$) all contained in a 375 MHz electrical bandwidth. The spectral shape and its fine structure are verified by an optical spectrum analyzer and direct spectroscopy, respectively. To assess the validity of the technique, we observe the absorption feature of acetylene at 1035.05 nm. The demonstrated quasi-real-time computational coherent averaging enlarges the toolbox for DCS measurement using free-running OFCs. Besides 750-MHz Yb:fiber combs, it could be also applied in real-time DCS using other high-repetition-rate combs, where tight phase locking of comb modes frequency and offset frequency is not applicable.

## 2. Principle

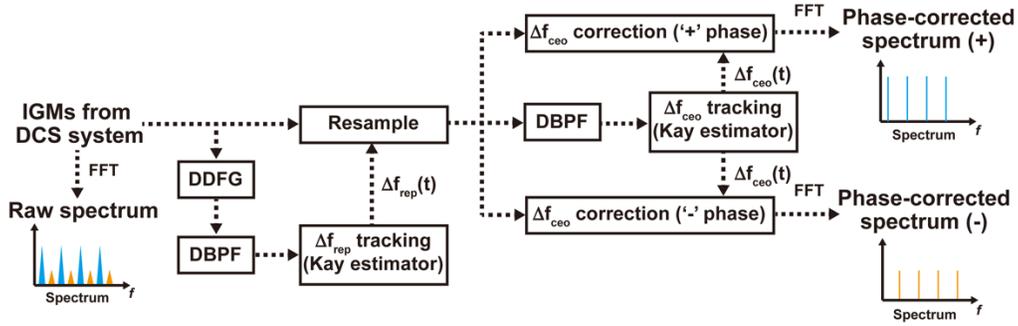

Fig. 1. Principle of computational coherent averaging dealing with aliased RF spectra. DDFG: digital difference frequency generation; DBPF: digital band-pass filter.

The general block diagram of the phase correction scheme used in our system is illustrated in Fig. 1. It mostly follows Ref. [26] with the modification to allow simultaneous in-phase and counter-phase offset frequency correction. The IGM sequence available from the photodetector is sampled by a fast digitizer. In our case, the IGM sequence consists of ~4600 individual IGMs produced every 217 ns, whose spectrum spans over three Nyquists zones (two sets of comb lines are aliased). The phase noise in the IGMs originates from fluctuations in the relative repetition rate difference ($\Delta f_{rep}$) and carrier-envelope frequency ($\Delta f_{ceo}$) between the two combs. For digital correction, two steps are required: time axis resampling and global offset frequency correction. First, the analytic signal, $\tilde{s}(t)$, of the sampled IGM sequence $s(t)$ is calculated through Hilbert transform. The digital difference frequency generation (DDFG) signal, $s_{DDFG}(t)$, is calculated through:

$$s_{DDFG}(t) = \text{Re}\{\tilde{s}(t)\}^2 + \text{Im}\{\tilde{s}(t)\}^2 \quad (1)$$

Note that the DDFG progress completely removes the effect of the offset frequency. Therefore, the DDFG spectrum contains only harmonics of the IGM's repetition rate in the frequency domain. A digital Kay frequency estimator is used to track the instantaneous frequency of the 64[th] harmonic (around 297 MHz) [34], which is filtered by a digital bandpass filter. Tracking high-order harmonics leads to the retrieval of $\Delta f_{rep}$ with higher precision than the fundamental. Instead of a digital frequency estimator, an IQ demodulator would also work for frequency tracking. The resampled time axis is then generated using the retrieved $\Delta f_{rep}$ according to:

$$t'(t) = \int_0^t \Delta f_{rep}(\tau)/\langle \Delta f_{rep}\rangle d\tau \qquad (2)$$

Where $t$ and $t'$ corresponds to the original time axis and the resampled time axis, respectively. $\Delta f_{rep}(\tau)$ is the tracked repetition rate of the IGM sequence, $\langle \Delta f_{rep}\rangle$ is its expected value. The resampled IGMs, $S(t')$, are obtained through interpolation using the sampled IGMs, original time axis $t$, and resampled time axis $t'$.

After resampling, as required for the next step of global offset frequency correction, one RF comb mode at around 255 MHz is digitally filtered and frequency-tracked in a similar fashion. The instantaneous frequency of this (or any other) comb line carries information about offset frequency excursions. They are compensated using:

$$S_c(t') = \exp\left(-i2\pi \int_0^{t'} \langle \Delta f_{ceo}\rangle - \Delta f_{ceo}(\tau)d\tau\right)\cdot S(t') \qquad (3)$$

Where $S_c(t')$ is the sampled IGMs signal after overall frequency correction. $\Delta f_{ceo}(\tau)$ is the tracked overall offset frequency, $\langle \Delta f_{ceo}\rangle$ is its expected value. Besides these correction steps provided originally in Ref. [24], we calculate a counter-phase corrected signal (flip the sign of $\Delta f_{ceo}$). As a result, the aliased RF spectra can be easily distinguished and unfolded manually.

## 3. Experimental setup and results

### 3.1 Offline phase correction

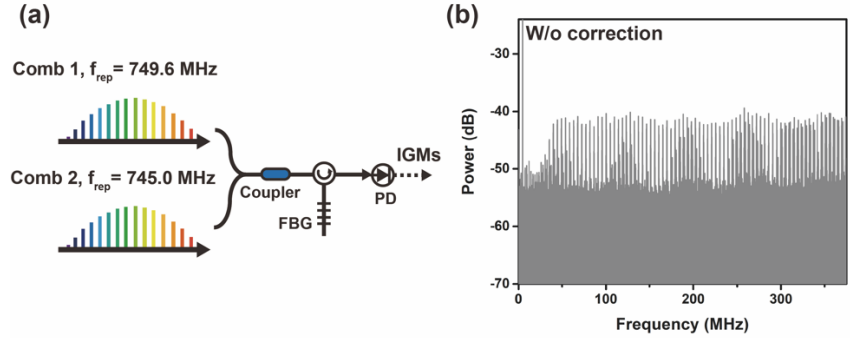

Fig.2 (a) Experimental setup for offline phase correction. FBG: fiber Bragg grating; PD: photodiode. (b) Spectra of IGM sequence before offline phase correction.

Two free-running Yb:fiber mode-locked lasers with $f_{rep,\,1}$=749.6 MHz and $f_{rep,\,2}$=745.0 MHz are used [35], as shown in Fig. 2(a). The pulse trains are first combined by a fiber coupler, and next guided through a fiber circulator connected to a reflective Bragg grating (FBG, center wavelength at 1030 nm, ~0.3 nm bandwidth). Finally, the filtered pulses are detected by a high-speed low-noise photodetector (Newfocus, 1611). The detected IGM sequence is sampled using a 14-bit digitizer (PXI-5164, National Instruments) with a 1 GS/s sampling rate. To reject additional replicas of the dual-comb signal above the digitizer's Nyquist frequency (500 MHz), we rely on the built-in low-pass filter with a 400 MHz cut-off frequency. Although the down-converted dual-comb spectrum clearly shows discrete comb modes in the RF domain, as seen in Fig. 2 (b), the RF comb modes drift due to the excess relative phase noise between the two free-running combs. This results in RF comb modes with poor visibility with a typical linewidth of 200 kHz. In addition to broadened lines, a moiré-like pattern with a scalloped envelope appears in the spectrum, which arises due to the multi-zone aliasing. The Nyquist range of this DCS measurement is around 0.2 nm (~60 GHz, ~ 80 comb lines), which causes overlap of three sets of RF spectra within the frequency range of DC to $f_{rep}/2$.

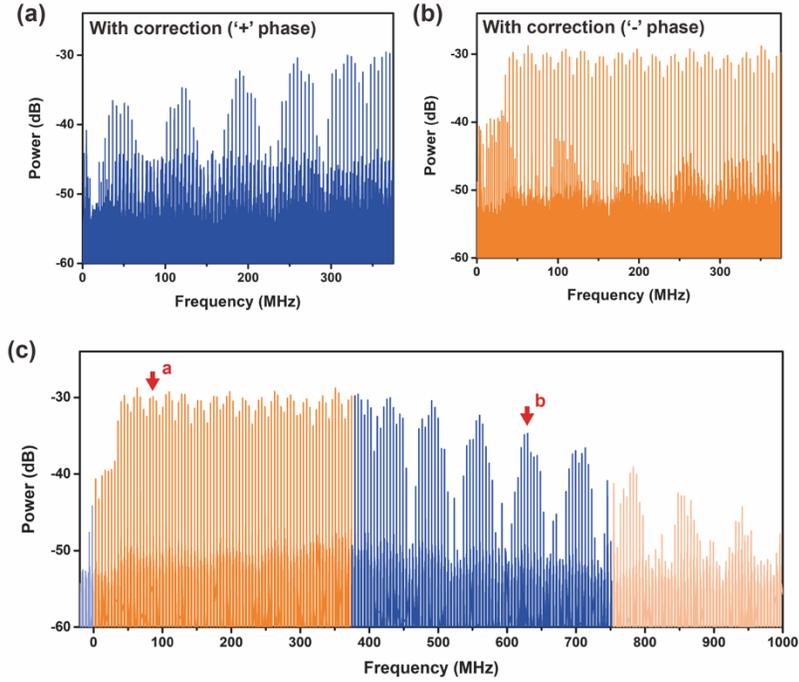

Fig.3 RF spectra following offline phase correction in phase '+' (a) and counter phase '−' (b), along with an unfolded RF spectrum (c). The red arrows point to the zoomed comb modes in Fig. 4 (a) and (b).

To verify the capability of phase correction with multi-zone aliasing, an offline experiment is first conducted. 1-ms long IGM sequences (about 4600 IGMs) are digitized, computationally resampled, and phase-corrected. The spectra of the IGM sequence with '+' (in-phase) and '−' (counter phase) correction are shown in Fig. 3 (a) and (b), respectively. The RF comb modes that correspond to different wavelength ranges in the optical domain are narrowed. After correction, the overlapped RF spectra are unfolded manually to recover the unaliased spectrum, as shown in Fig. 3 (c). In Fig. 4 (a) and (b), we present zoomed comb modes at around 86.4 MHz and 628.8 MHz, respectively. Compared with the free-running RF comb modes (represented as gray curves), the corrected comb linewidth is narrowed from 200 kHz to 1 kHz, while the line-to-floor ratio (LFR) of the RF comb modes increases 13 dB in power. It should be noted that the linewidth already reaches the theoretical limit, which is the inverse of the sampling time. Correction of data acquired over 10 ms (Fig. 4 (c)), and 21 ms (Fig. 4 (d)) also yields Fourier-limited linewidths of ~ 100 Hz and ~ 50 Hz, respectively.

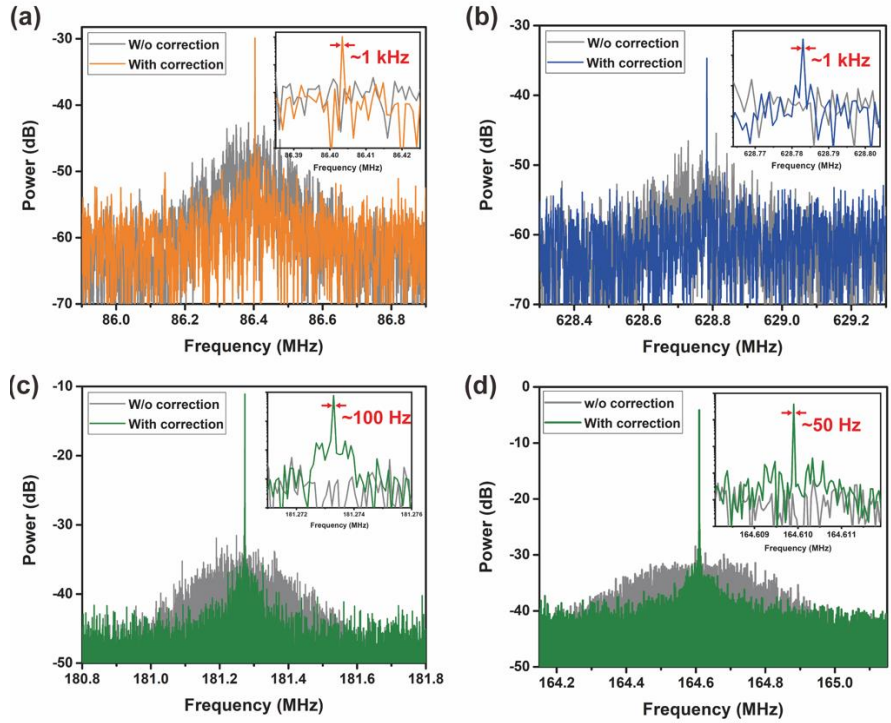

Fig. 4 Representative RF comb modes in the IGM's spectrum with offline phase correction. The comb mode was at 86.4 MHz (a) and 628.8 MHz (b) in the 1 ms data set. (c) The comb mode at 181.3 MHz in the 10 ms data set. (d) The comb mode at 164.6 MHz in the 21 ms data set. The insets show the zoom-in view of the comb modes.

### 3.2 Quasi-real-time digital correction

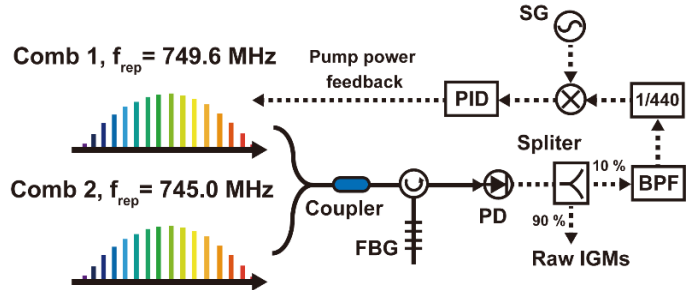

Fig. 5. Experimental setup of quasi-real-time DCS measurement. FBG: fiber Bragg grating; PD: photodiode; BPF: band-pass filter; SG: signal generator; PID: proportional-integral-derivative servo controller.

In the case of offline phase correction, the free-running RF comb lines would easily drift out of the range of the digital band-pass filter on a long-term scale. This prevents us from achieving real-time phase correction and further power averaging RF spectra. To address this issue, slow phase locking of the RF comb lines is implemented in the DCS system. As shown in Fig. 5, a microwave power splitter couples 10% of the electrical power for phase locking. One of the RF comb modes is filtered with an RF band-pass filter centered at 70 MHz. It is next amplified to > 0 dBm, frequency-divided, and finally mixed with an RF reference from a signal generator. The generated error signal is sent into a PID servo controller, which generates a correction signal to modulate the pump power of Yb Comb 1. The locking bandwidth of this simple yet robust phase-locked loop (PLL) is in the kHz range. Therefore it does not narrow the RF comb mode. It merely ensures that the locked RF comb line always stays within the filter bandwidth and that the aliased comb lines do not spectrally overlap (remain interleaved).

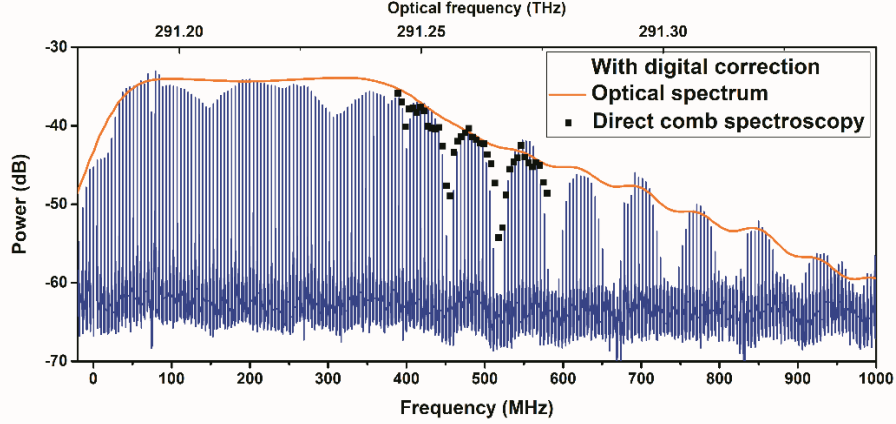

Fig. 6. Spectra of IGM sequence in quasi-real-time DCS measurement.

After coarse phase locking, the IGM sequence is sampled at a 1 GHz sampling rate for 1 ms, and then phase-corrected for ~1.5 s. In this demonstration, the routine is repeated for 20 times (see Visualization 1 in the supplementary material) leading to a total acquisition and processing time of ~30 s. Owing to the limited processing capability of the computer (digitizer's hardware resources), the system cannot provide continuous online phase correction, yet it still provides new coherently averaged spectra in an acceptable time of fewer than 2 seconds. Although this mode of operation can be termed quasi-real-time, the speed limit can be mitigated in future FPGA-based implementations instead of software-based filters and demodulators. Just like in the offline case, all the RF comb modes are easily distinguishable after phase correction and power averaging of 20 spectra, as shown in the blue curve in Fig. 6. Similar to the offline case, the aliased RF lines are manually unfolded to generate the DCS spectrum spanning 3 Nyquist zones. Note that if without phase-locking and correction, the RF comb modes would be dispersed and spread across a wide range of frequencies with a low SNR. The LFR of RF comb modes at 199.8 MHz and 849.8 MHz are around 30 dB and 11 dB in power after correction. The SNR in DCS measurement is determined by the baseline fluctuation [36-39]. Since it is comb-line-resolved DCS, the SNRs are calculated by the amplitude fluctuation of the comb modes. We obtained 10.4 and 7.3 in 1 ms for a comb mode at 79.2 MHz and 282.8 MHz, respectively, which fall into a typical range in comb-line-resoled DCS systems [14,39].

To confirm that the scalloped, unfolded part of the spectrum above 400 MHz is not a measurement artifact resulting from comb filtering (superposition of time-delayed signals) or an aliasing issue, we directly measure the filtered optical spectrum using an optical spectrum analyzer (Yokogawa, AQ6370D). The general shape of the RF spectrum shows agreement with the optical spectrum (as shown in the orange curve in Fig. 6). We also verify the fine structures (spectral dips) in the obtained RF spectrum using the direct spectroscopy technique [40]. The filtered comb modes from Comb 1 are beaten with a narrow-linewidth CW laser at 1030 nm (Connet, CoSF-D-YB-B-LP, < 20 kHz typical linewidth). The heterodyne products are detected using a 40 GHz photodiode (DSC10H, Discovery) and next rearranged. Their optical frequencies are then calibrated, as shown in the black squares in Fig. 6. The two spectral dips from this measurement are consistent with those in the measured dual-comb spectrum at 291.256 THz and 291.266 THz. We attribute the regularly-spaced dips to the reflection properties of the FBG, which includes multiple quasi-periodic sidelobes.

*3.3 Absorption measurement of acetylene gas cell*

To assess the suitability of the developed DCS system for molecular spectroscopy, we measure an absorption feature of acetylene at atmospheric pressure. The FBG is replaced with one that matches the wavelength range of the target absorption line (1035.01 nm center wavelength, 0.24 nm bandwidth). 50 IGM reference sequences (without the acetylene) are sampled for 1 ms and phase-corrected in quasi-real-time, as illustrated in Section 3.2. The amplitude of the 50 corrected RF spectra is then averaged. Next, a fiber-type acetylene gaseous cell (Wavelength References, 740 torr, 16.5-cm long) is inserted after the combination of two combs. The insertion loss is 5 dB. To obtain the gaseous cell transmission spectrum, we calculate the ratio of the RF spectra (in amplitude) with and without the sample. The position of absorption feature located at 1035.05 nm is consistent with HITRAN database, as shown in Fig. 7 (a).

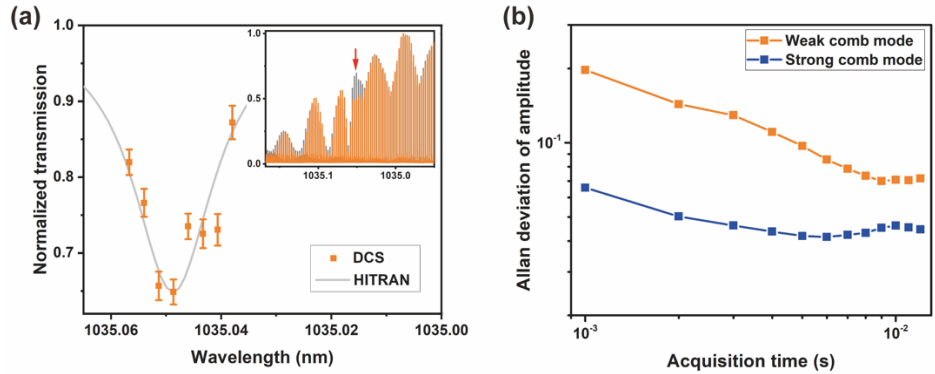

Fig. 7. (a) Absorption spectrum of acetylene gas cell using quasi-real-time DCS measurement. The error bar is calculated from the standard deviation of the comb modes' amplitude with gaseous cell divided by square root of number of measurements. Inset: normalized amplitude versus wavelength in nanometer of resolved RF comb modes with (orange) and without (gray) gaseous cell. The red arrow points to the absorbed comb modes. (b) Normalized Allan deviation of amplitude of two representative RF comb modes at 268.0 MHz (strong comb mode) and 556.1 MHz (weak comb mode).

The demonstrated quasi-real-time phase correction enables long-term power averaging of multi-corrected RF spectra, which in turn enhances the SNR of the RF comb modes. This is essential for precise spectroscopic applications. We performed an Allan deviation analysis of two representative RF comb modes' amplitude with different power located around 268.0 MHz (strong comb mode) and 556.1 MHz (weak comb mode) in the 50 corrected RF spectra obtained from absorption measurement. The result shown in Fig. 7 (b) reveals that the algorithm is capable of reaching lower uncertainty of the amplitude estimate at prolonged integration time. For both the strong and weak comb mode, percentage-level precision becomes accessible in > 10 ms acquisition time. The SNR in DCS is 10.02 in 1 ms for the comb mode at 268.0 MHz. One concern is that how the stability of the free-running comb modes in the fiber laser contributes to the resolution of our spectroscopy measurement. To investigate this, we characterize the stability of the optical comb modes at different time scale. In 1 ms (> 500 s) time scale, the stability reaches to $1.44\times10^{-9}$ ($2.66\times10^{-7}$), corresponding to spectroscopy resolution of 0.0015 pm (0.23 pm), which has merely influence for determination of most gaseous absorption line position.

## 4. Summary

In summary, we demonstrate quasi-real-time DCS based on two 750-MHz Yb:fiber combs with coarse relative phase locking and the computational coherent averaging technique to compensate for the phase noise in the IGM sequence. After phase correction, the linewidth of the RF comb mode in the spectrum of the IGM sequence in 1 ms sampling time is reduced from

200 kHz to ~ 1 kHz with 13-dB LFR enhancement and can be further narrowed to ~100 Hz and ~ 10 Hz in 10-ms and 21-ms time scales, respectively. Applying simultaneous in-phase and counter phase correction leads to distinguishable subsets of aliased RF comb lines spanning 3 Nyquist zones. To achieve quasi-real-time phase correction, one of the RF comb modes at 70 MHz is phase-locked to an RF reference through pump power feedback. The simple yet robust phase-locking scheme prevents the comb modes from drifting, which is a significant prerequisite for RF spectral interleaving when the span of DCS lines exceeds the Nyquist range. In this demonstration, 1-ms-long acquisitions are phase-corrected in an acquisition software within 1.5 seconds. Next, 20 phase-corrected spectra are power-averaged. The recovered RF spectrum's general shape and fine structure are verified by an optical spectrum analyzer and direct spectroscopy, respectively.

In general, high repetition dual-comb system can achieve broad measurement frequency range. However, in some cases, the tunability of repetition rate is limited due to engineering issue that leads to large repetition rate difference between the two combs and narrower Nyquist range. Such a situation is commonly seen in high-repetition rate combs' spectroscopic application, e.g., microcombs, monolithic lasers [41]. Our present study provides an effective scheme to solve the multi-aliasing of RF comb modes due to the narrowed Nyquist bandwidth. Previously, other relevant methods have been reported, such as application of combs with a larger spacing, e.g. > 100 GHz spacing [11]. However, a high bandwidth photodetector is required if one wants to resolve such as more than 100 comb modes. The spectroscopy resolution is also limited at > 100 GHz, preventing from absorption profile measurement of some gaseous sample. Arbitrary detuning asynchronous optical sampling (ADASOPS) can also deal with repetition rates differ by a great amount, demonstrating pump-probe spectroscopy with kHz repetition rate laser system [42-43]. However this technique requires complexity in hardware such as coincidence detection and is difficult to apply in GHz laser system. Therefore, our proposed scheme is advantageous.

The demonstrated approach surely has its limitation. The number of aliased Nyquist zones is limited by the linewidth of the free-running heterodyne beats. In this case, we expect a maximum number of 4 to 5 aliased Nyquist range, corresponding to 320 to 400 RF comb modes. More free-running RF comb modes aliasing in one Nyquist zone will cause overlapping thus the coarse phase locking, frequency tracking and phase correction would not work perfectly. However, if the free-running relative linewidth is smaller, the applicable number of aliased Nyquist range is expanded. We observe the absorption of acetylene at 1035.05 nm when inserting a fiber gaseous sample. To achieve a better spectroscopy measurement, a baseline without the effect of artifacts of the optical filter (FBG) is required. Using a grating and pair of blades as an optical filter is a potential solution. It should be noted that the CoCoA algorithm offers lower computational complexity for systems with nearly-isolated MHz-spaced RF lines. In our case, RF comb lines can be easily identified and filtered using conventional demodulators, filters, etc., all compatible with modern digital signal processing (DSP) platforms. The full spectroscopic potential of the system will be explored with future wavelength conversion to the mid-infrared or THz region. The near-GHz resolution is highly attractive for studying molecular absorbers at lower pressures, while the sub-microsecond resolution should be sufficient for time-resolved studies.

**Funding.** This work was supported by Japan Society for the Promotion of Science (JP21H05014); Horizon 2020 Framework Programme (H2020) (101027721).

**Acknowledgments.** We thank B. Xu, H. Yasui, and H. Ishii of University of Electro-Communications, and Z. Zhang of Peking University for development of the Yb comb used in this study. We also thank H. Song in SLAC National Accelerator Laboratory for useful discussion. L. A. Sterczewski acknowledges funding from the European Union's Horizon 2020 research and innovation programme under the Marie Skłodowska-Curie grant agreement No 101027721.

**Disclosures.** The authors declare no conflicts of interest.

**Data availability.** Data underlying the results presented in this paper are not publicly available at this time but may be obtained from the authors upon reasonable request.

**Supplemental document.** See Visualization 1 for a recorded video of quasi-real-time DCS measurement.